\documentclass[%
 reprint,
 amsmath,amssymb,
 aps,
]{revtex4-2}

\usepackage{graphicx}% Include figure files
\usepackage{dcolumn}% Align table columns on decimal point
\usepackage{bm}% bold math
\usepackage{gensymb}

\begin{document}

\title{Pulsed Dual-axis Alkali-metal-noble-gas Comagnetometer}

\author{J. Wang}\thanks{Electronic address: jingyaow@princeton.edu}%Jingyao Wang
\author{J. Lee}%Junyi Lee
\author{H. Loughlin}%Hudson Loughlin
\author{M. Hedges}%Morgan Hedges
\author{M. V. Romalis}\thanks{Electronic address: romalis@princeton.edu}%Michael V. Romalis
\affiliation{%
Department of Physics, Princeton University, Princeton, New Jersey 08544, USA
}%

\date{\today}

\begin{abstract}
Alkali-metal-noble-gas comagnetometers are precision probes well-suited for tests of fundamental physics and inertial rotation sensing, combining the high sensitivity of the spin-exchange-relaxation free (SERF) magnetometers with inherent suppression of magnetic field noise. Past versions of the device utilizing continuous-wave optical pumping are sensitive to rotation and anomalous spin couplings along a single axis perpendicular to the plane spanned by the orthogonal pump and probe laser beams. These devices are susceptible to light shifts in the alkali atoms, and to power and beam pointing fluctuations of both the probe and pump lasers, the latter of which is a dominant source of \(1/f\) noise. In this work, we model and implement an approach to alkali-metal-noble-gas comagnetometers using pulsed optical pumping. After each pump laser pulse, an off-resonance probe beam measures the precession of noble-gas-spins-coupled alkali spins via optical rotation in the dark, thus eliminating effects from pump laser light shift and power fluctuations. Performing non-linear fitting on the sinusoidal transient signal with a proper phase enables independent and simultaneous measurement of signals along two orthogonal axes in the plane perpendicular to the pump beam. Effects from beam pointing fluctuations of the probe beam in the pump-probe plane are fundamentally eliminated, and signal response to pump beam pointing fluctuations is suppressed by compensation from noble-gas nuclear spins.
\end{abstract}


\maketitle

Searches for new physics at low energy often involve precision measurement of anomalous spin-dependent interactions \cite{safronova18}. Many comagnetometer systems have been demonstrated to be well-suited for this purpose \cite{terrano21}, relying on the differentiation of systematic effects from signal of interest made possible by measuring common systematic effects with one spin ensemble, therefore revealing weak signals in the other. Different interactions under investigation command different combinations of spin species, for instance, two electron-spin species for electron electric dipole moment (EDM) searches \cite{weisskopf68,Chupp19}, two nuclear-spin species for \(^{129}\)Xe EDM measurements \cite{Rosenberry01} and tests of Lorentz invariance \cite{lamoreaux86} and CPT symmetry \cite{bear00}, or a pair of electron-spin and nuclear-spin species for Lorentz invariance test \cite{berglund95} and axion-like coupling searches \cite{youdin96}. The two ensembles may also consist of the same species in a spatially non-overlapping configuration \cite{lamoreaux87}, in a Bose-Einstein condensate with two distinct hyperfine manifolds \cite{Gomez20}, or in an ensemble of molecular species where different nuclear spins in the same molecule are compared \cite{wu18}.
\\\indent In addition to reduced systematic effects with differential output from two spin ensembles, a comagnetometer's sensitivity can be further boosted by utilizing a highly sensitive magnetic field readout system. The state of the art of such systems is set by spin-exchange relaxation free (SERF) atomic magnetometers \cite{allred02,kominis03,dang10}, which take advantage of the disappearance of spin-exchange broadening at high alkali-metal vapor density and in a near-zero magnetic field \cite{happer73,happer77}. This work focuses on one type of alkali-metal-noble-gas comagnetometer with the SERF magnetometer incorporated into its setup \cite{kornack02}. The alkali-metal atoms both polarize noble-gas atoms via spin-exchange optical pumping \cite{walker97,bouchiat60} and probe their precession via spin-exchange interaction \cite{schaefer89}, while the noble-gas atoms suppress sensitivity to magnetic field drifts, gradients, and transients \cite{kornackThesis,kornack05}, but retain sensitivity to anomalous magnetic-like fields. Such self-compensating comagnetometers have found application in inertial rotation sensing \cite{kornack05, li16}, as a quantum interface \cite{QI}, as well as in fundamental physics, including tests of Lorentz and CPT violation \cite{smiciklas11}, searches for anomalous spin-mass \cite{lee18} and spin-spin \cite{almasi20} interactions, and direct detection of dark matter \cite{lee23}.
\\\indent Despite having undergone a string of revisions, past versions of the self-compensating comagnetometer still have several limitations. In the initial K-\(^3\)He comagnetometer, unlike how magnetic fields are suppressed by coupled spin dynamics, light shifts \cite{mathur68} and optical misalignment can be zeroed with an elaborate zeroing procedure \cite{kornackThesis}, but not fundamentally suppressed, since they couple only to alkali spins. Their drifts over time contribute to \(1/f\) noise, whose presence requires all fundamental physics searches to adopt some form of modulation, either of the source or of the whole experiment. A subsequent K-\(^3\)He comagnetometer improved low-frequency performance and achieved the lowest spin energy resolution, corresponding to a magnetic field value less than an aT (10\(^{-14}\) G), by quickly reversing the source spin and modifying pump optics and heating scheme to minimize perturbation of the experiment \cite{vasilakis09}. The next version further shortened optical path length, enclosed all optics in an evacuated bell jar, and periodically reversed the whole setup, which resulted in better \(1/f\) noise suppression \cite{brown10}. In spite of the improvements, fundamental suppression of light shifts and optical misalignment was still lacking. Later adoption of hybrid spin-exchange optical pumping \cite{babcock03}, where optically-pumped low-density Rb vapor transfers polarization to collocating high-density K vapor via spin-exchange collisions \cite{lee18}, reduced polarization gradient in the latter and thus optimized sensitivity, but also introduced a light shift that can no longer be zeroed due to the circularly-polarized pump laser detuned from the K's resonance frequencies. Replacing K with Cs to hybrid pump Rb \cite{almasi20,jiang18} in a Rb-\(^{21}\)Ne comagnetometer ameliorated the problem by increasing the detuning, and separate studies reduced this light shift by compensation with either light shift in K \cite{chen16} or magnetic field \cite{jiang17} in the K-Rb-\(^{21}\)Ne system, or exploited it to sense dual-axis rotations \cite{li16}, whilst the traditional steady-state DC-signal comagnetometers are sensitive to a single axis. 
\\\indent Alternative operating schemes have been investigated to realize dual-axis sensitivity, including introducing high-frequency modulation of either the longitudinal magnetic field along the pump beam \cite{jiang18}, or two orthogonal transverse magnetic fields \cite{jiang22}, each combined with synchronous demodulation. Both schemes demonstrated improved \(1/f\) noise suppression compared to the DC-mode of operation, although their short-term sensitivities were worse due to magnetic noise introduced by the modulations. Noticeably the latter adopted a single-beam configuration, showing promise for miniaturization.
\\\indent In this work, we describe a dual-axis \(^{87}\)Rb-\(^{21}\)Ne comagnetometer utilizing pulsed optical pumping, which introduces minimal noise compared to the magnetic field modulations mentioned above, as the noble-gas atoms are unperturbed by optical modulation. By turning off the pump laser during measurement, pump laser light shift and power fluctuations afflicting DC-mode comagnetometers can be fundamentally eliminated or suppressed. Performing non-linear fitting on the oscillating transient signal enables separation of beam misalignment from the real signal due to rotation and anomalous spin couplings, thereby eliminating one dominant source of \(1/f\) noise from beam pointing fluctuations. Here we present results from both experiment and Bloch equation model to demonstrate the above advantages of the pulsed comagnetometer. Experimentally, the elimination of \(1/f\) noise is not yet realized, and we discuss possible limitations of our setup. 

\section{\label{sec:level1}Overview}

\begin{figure}
\includegraphics[width=8.5cm, keepaspectratio]{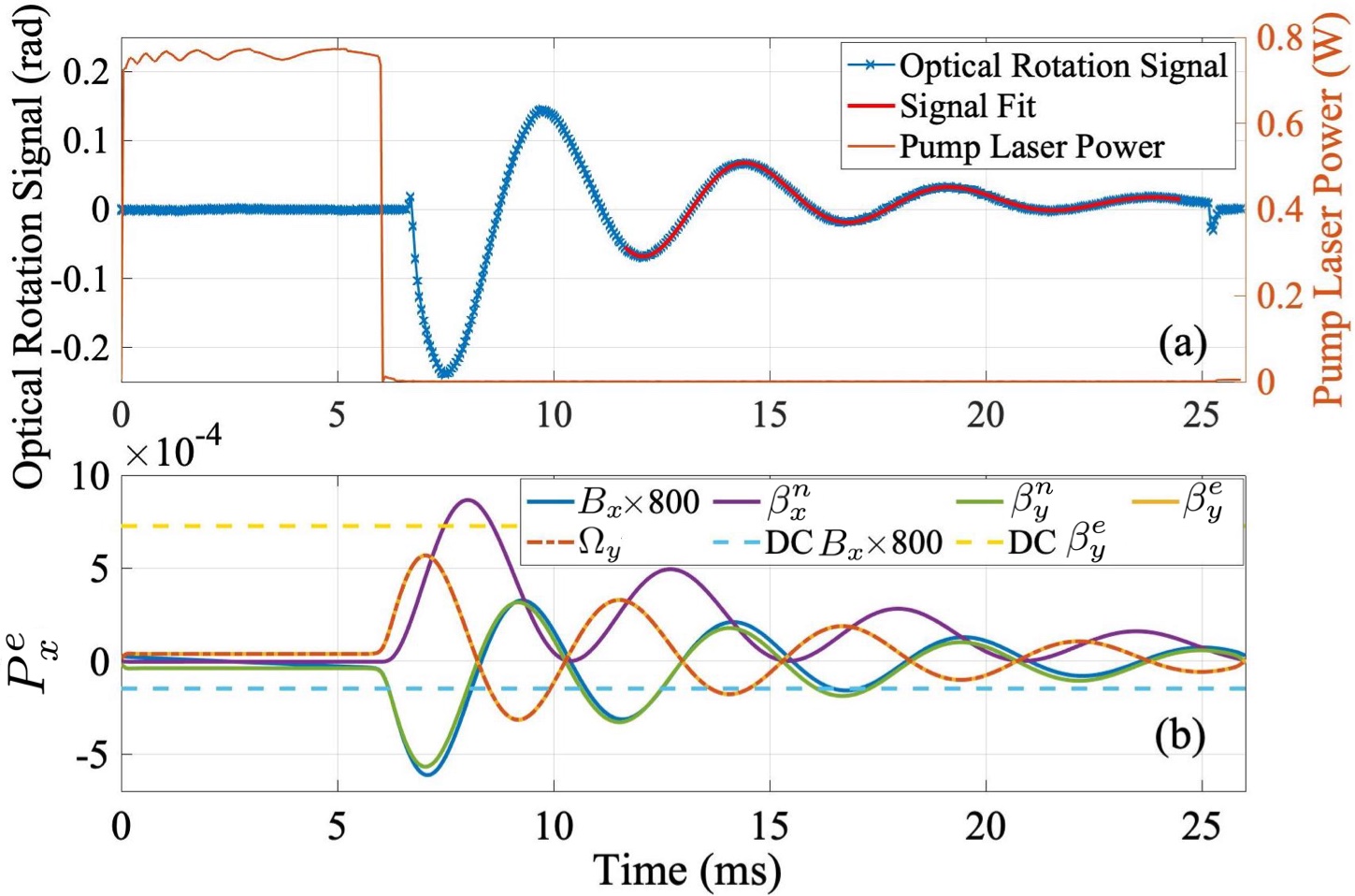}
\caption{\label{fig:RawSignal} (a) Left scale: Optical rotation signal for one pump period and its fit in the form of Eqn.~\ref{eqn:fittingForm}. An analog switch blocks photodiode signal during pumping pulse to avoid overloading the lock-in amplifier. The fit starts at 11 ms to avoid the faster-drifting part of oscillation frequency due to polarization-dependent slowing down factor \(Q(P^e)\). Right scale: Pump laser power. Power fluctuations during the pulse result from quick ramp-up of amplifier current. (b) Numerically simulated electron spin \(\hat{x}\)-polarization at the compensation point \(B_z=B^c=-(B^n+B^e)\). The various curves represent signal response to \(10^{-7}\) G of \(B_x\), \(\beta^n_{x}\), \(\beta^n_{y}\), or \(\beta^e_{y}\) field, and to a rotation \(\Omega_y=0.2\) mrad/s, corresponding to a \(10^{-7}\) G effective field. The \(\beta\) symbols stand for anomalous magnetic-field-like spin interactions that couple only to the spin species in their superscript. Response to \(B_x\) is enlarged by a factor of 800. Responses to \(\beta^n_{x}\) and to \(\beta^n_{y}\) are \(\pi/2\) out of phase, and responses to \(\beta^e_{y}\) and to \(\beta^n_{y}\) are in-phase with opposite signs. Two numerically simulated curves for DC-mode in response to \(10^{-7}\) G of \(B_x\) and \(\beta^e_{y}\) are included for comparison.}
\end{figure}

For the pulsed comagnetometer, we chose \(^{87}\)Rb's electron spins and \(^{21}\)Ne's nuclear spins as the two sensing ensembles due to \(^{21}\)Ne's smaller gyromagnetic ratio compared to \(^{3}\)He, which improves fundamental sensitivity, and Rb's larger spin-exchange cross section with \(^{21}\)Ne compared to K \cite{Ghosh10}. Enriched \(^{87}\)Rb is favored over Rb in natural abundance for simplified spin dynamics with only one type of nuclear spin (\(I=3/2\)). Optical pumping polarizes \(^{87}\)Rb atoms, which then polarize \(^{21}\)Ne via spin-exchange collisions. The circularly polarized pump laser along  \(\hat{z}\)-axis is tuned to Rb's D1 resonance, and cycles between 6 ms of pumping time, where \(^{87}\)Rb builds up high longitudinal polarization, and 20 ms of dark time, where the polarization precesses and decays. A detuned probe laser along  \(\hat{x}\)-axis continuously monitors the spins' \(\hat{x}\)-projection via optical rotation. In the presence of transverse coupling, i.e. spin interaction in the plane orthogonal to \(\hat{z}\)-axis, to either \(^{87}\)Rb, \(^{21}\)Ne, or both, the motion of \(^{87}\)Rb's electron spins is a transient hybrid response coupled to \(^{21}\)Ne nuclear spins via spin-exchange, resembling Larmor precession, as shown in Fig.~\ref{fig:RawSignal}(a).  The spin-exchange coupling between the two species due to Fermi-contact interaction can be represented as effective fields \(B^n = \lambda M^{n}\) and \(B^e = \lambda M^{e}\) acting on one another, where \(M^{e}\) and \(M^{n}\) are the magnetizations of electron and nuclear spins. For a spherical cell  \(\lambda = 8\pi \kappa_{0}/3\), and the enhancement factor \(\kappa_{0} = 35.7\) for Rb-Ne \cite{Ghosh10}. Typically \(B^n\) is a few mGs and \(B^e\) a few hundred \(\mu\)Gs. Information on transverse spin interactions is contained in the amplitude of the decaying oscillations, which is extracted by fitting the signal to 
\begin{align}\label{eqn:fittingForm}
S(t)=&[A\text{sin}(2\pi ft+\phi_0)+B\text{cos}(2\pi ft+\phi_0)]e^{-t/T_2}\nonumber\\
&+Ce^{(-t/T_1)}+D, 
\end{align}
where \(T_1\) and \(T_2\) denote longitudinal and transverse relaxation times respectively, and \(\phi_0\) is a fixed parameter determined from a zeroing procedure described later. 

\begin{figure}
\includegraphics[width=8.5cm, keepaspectratio]{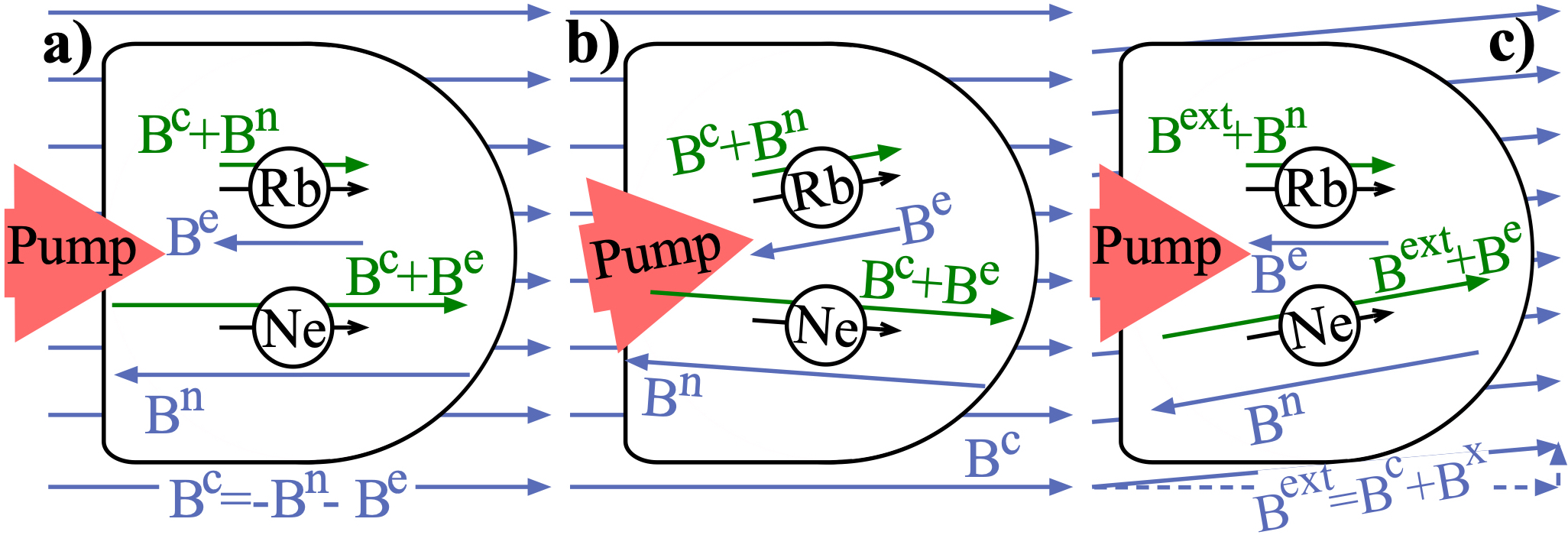}
\caption{\label{fig:SuppDiagram} Basic operation of the comagnetometer. (a) Electron spins and noble-gas nuclear spins align with the applied bias field \(B^c=-(B^n+B^e)\). Each species generates an effective field felt by the other species indicated by the blue arrows, and feels a field indicated by the green arrows. (b) Deflected pump laser polarizes electron spins along the tilted direction, generating a tilted \(B^e\). Noble-gas nuclear spins feel a tilted field \(B^e+B^c\) and adiabatically follow it, generating a tilted \(B^n\). Electron spins feel the field \(B^n+B^c\), which largely aligns with tilted pump laser direction. When the pump laser is turned off the electron spins' precession is suppressed as the spins are aligned with the field they feel, and the probe laser picks up an exponential decay due to \(T_1\)-relaxation if the deflection results in non-orthogonal pump and probe directions. (c) Noble gas nuclear spins adiabatically follow the change in external field, generating a tilted \(B^n\), whose transverse component cancels the transverse component of external field for electron spins. When the pump laser is turned off the electron spins' precession is suppressed as the spins are aligned with the field they feel, and their projection along the probe laser remains zero.}
\end{figure}

As summarized in Tab.~\ref{tab:table1}, the pulsed comagnetometer exhibits multiple advantages compared to past versions of the self-compensating comagnetometer operating in DC-mode, while maintaining their key feature of suppressed response to magnetic fields, as shown in Fig.~\ref{fig:SuppDiagram}(c).
First, the pulsed comagnetometer has dual-axis sensitivity in the transverse plane because any transverse coupling induces electron spin precession, whose \(\hat{x}\)-projection exhibits oscillatory behavior. By comparison, in the DC-mode transverse \(\hat{x}\)-couplings are not detected because they tilt steady-state electron spins into \(\hat{y}\)-axis, causing no change in their \(\hat{x}\)-projection. Second, any \(\hat{z}\)-axis projection of the probe laser shows up as an exponential decay in the signal, since the probe picks up a portion of the longitudinal electron polarization with \(T_1\)-relaxation. This can be easily differentiated from oscillations, whereas such misalignment shows up as a DC-offset in DC-mode, indistinguishable from the transverse couplings of interest. Third, light shift from the pump laser is fundamentally eliminated in the pulsed scheme, as the pump laser is not present when measurements are taken. Fourth, the pulsed comagnetometer has suppressed sensitivity to pump laser power fluctuations at sufficiently high pumping rate \(R_p\). While the DC-mode magnetometer's sensitivity is maximized at 50\(\%\) electron spin polarization, the pulsed case's sensitivity increases with equilibrium electron spin polarization when the pump laser is on, and its \(R_p\)-dependence asymptotically approaches zero with increasing \(R_p\). Aiming for maximum electron spin polarization during the pulse also means the pump beam burns through the cell causing no polarization gradient (with a well-collimated pump beam illuminating all spins). Thus hybrid pumping is not necessary, and spin dynamics can be simplified. Finally, re-orientation of noble-gas nuclear spins in response to the tilted electron spin polarization due to pump beam deflections suppresses oscillation amplitude of the electron spins during dark time, as shown in Fig.~\ref{fig:SuppDiagram}(b). In DC-mode a tilted pump beam non-orthogonal to probe beam results in a DC-offset indistinguishable from the signal of interest.

\begin{table}
\caption{\label{tab:table1}Comparison between DC-mode and pulsed-mode of comagnetometer operation.}
\begin{ruledtabular}
\begin{tabular}{  m{5.5cm}  m{1.5cm} m{2 cm}  } %{lcr7
Comagnetometer Feature &DC-mode &Pulsed\\
\hline
Magnetic Field Suppression & Yes & Yes\\
Sensitive Axis & \(\hat{y}\)-axis & \(\hat{x}\)- and \(\hat{y}\)-\\
Probe \(\hat{z}\)-misalignment Differentiation & No & Yes\\
Pump Light Shift Elimination & No & Yes\\
Pump Power Fluctuation Sensitivity & Yes & No \footnote{At high \(R_p\).}\\
Pump Beam Deflection Suppression & No & Yes \\
\end{tabular}
\end{ruledtabular}
\end{table}

At the same time, we find that some non-magnetic interactions cannot be eliminated. For example, probe laser residual light shift creates a non-magnetic interaction that couples only to electron spins and can create a systematic signal, although this is not a limiting factor in our experiment. We also find that non-uniformity of the pump laser direction in the cell creates complicated dynamics that are likely responsible for \(1/f\) noise in our system.  

\section{\label{sec:level1} Experimental Apparatus}

\begin{figure}
\includegraphics[width=8.6cm, keepaspectratio]{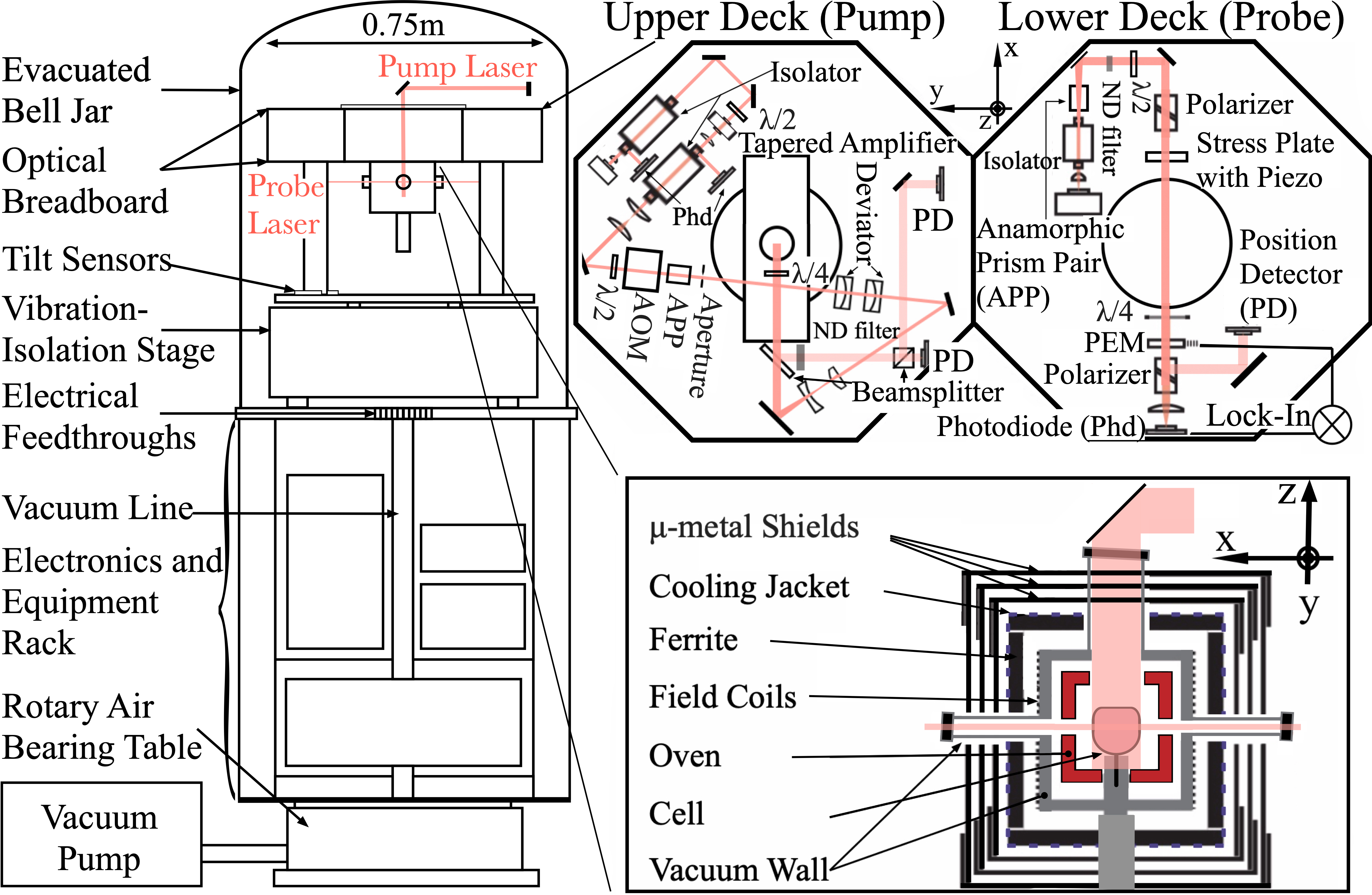}
\caption{\label{fig:Setup} Schematic diagram of the apparatus. Cell size is exaggerated. Probe optics are mounted upside down on the lower deck. Windows for optical access into the inner vacuum chamber, which is pumped to higher vacuum than inside the bell jar for thermal stability, are slightly tilted to prevent the formation of standing waves. The pump seed laser, tapered amplifier, probe laser, and case of AOM are water-cooled to 18 $^\circ$C on the same line.}
\end{figure}

Our experimental apparatus is shown in Fig.~\ref{fig:Setup}, modified from the one used in \cite{smiciklas11}. The pump beam is generated by a master-oscillator tapered power-amplifier setup with a DBR diode as seed laser, and then expanded to illuminate the whole cell, providing close to 800 mW of pumping power. The pulses are switched on-and-off with an acoustic-optic modulator (AOM) in combination with laser amplifier current modulation to minimize light leakage during dark time. Two beam direction deviators are placed at specific distances, determined with ABCD-matrix calculations, in relation to the lenses for beam-expansion and to each other, such that each separately controls the pump beam's tilt and translation at the cell. Probe laser optics follow a standard polarimetry set up for optical rotation using a photoelastic modulator (PEM), with an added stress plate for cancelling any stray birefringence. Photodiode output is demodulated with a lock-in amplifier and then recorded by a computer that performs non-linear fitting of the signal in real time. The probe beam is linearly polarized, with a diameter of 2 mm and a power of 2 mW.  Multiple position detectors and photodiodes monitor laser beam power, position, and angle. All optics are enclosed in a bell jar evacuated to low vacuum to eliminate noise from air currents. The bell jar is wrapped in insulation material to avoid coupling to environmental temperature changes. The whole experiment sits on a rotary air bearing table so its orientation can be rotated around the vertical axis.

\begin{figure}
\includegraphics[width=5.5 cm, keepaspectratio]{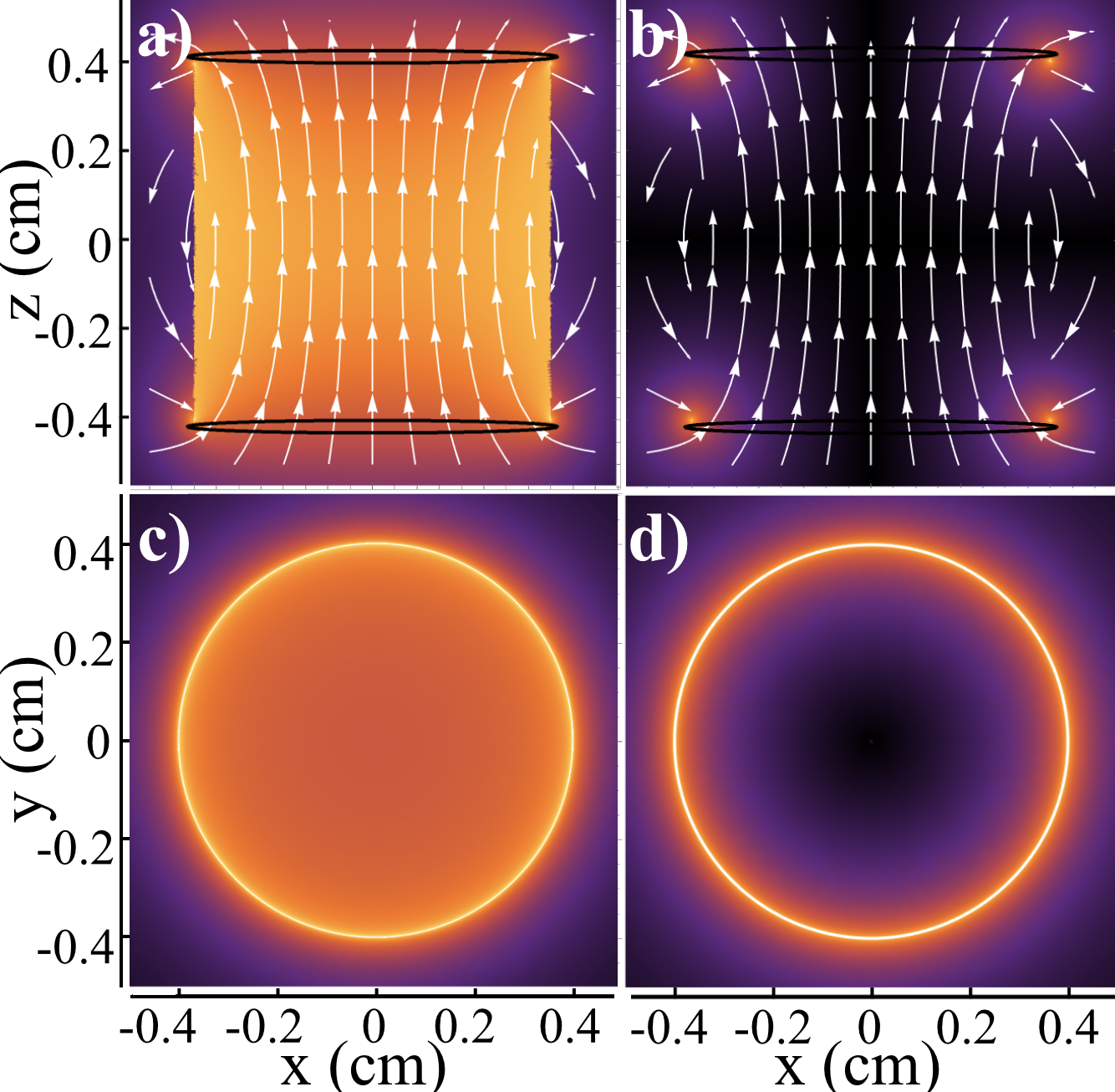}
\caption{\label{fig:NeDipolar} Model for Ne's dipolar field strength inside the cell as magnetic field strength for a cylinder with uniform longitudinal magnetization pointing towards positive \(\hat{z}\)- direction. The cylinder's length and diameter are both 0.8 cm. Its center sits at the origin. Lighter color signifies larger field magnitude. Ovals with black outline in (a) and (b) represent top and bottom surfaces of the cylinder. (a) Magnetic flux lines on top of total field strength in the plane \(y=0\). (b) Magnetic flux lines on top of transverse field strength in the plane \(y=0\). (c) Total field strength in the plane \(z=0.4\) cm. (d) Transverse field strength in the plane \(z=0.4\) cm.}
\end{figure}

The comagnetometer cell contains enriched \(^{87}\)Rb, around 5 atmosphere of \(^{21}\)Ne at room temperature, and 200 torr of N\(_{2}\) for quenching. It is heated to 170 \(^\circ\)C, at which Rb's number density is about 2.6\(\times10^{14}\) cm\(^{-3}\). The shape of the cell is a compromise between minimizing magnetic field gradients from dipolar fields created by Ne polarization and creating a uniform illumination of the cell by the pump laser.
The hand-blown glass cell is roughly a cylinder of diameter 0.8 cm and height 0.8 cm. Its axis is parallel to the pump beam, with a flat entrance window but a modified tapered shape where the pump beam exits, as illustrated in both Fig.~\ref{fig:SuppDiagram} and Fig.~\ref{fig:Setup}. The flat entrance window helps minimize pump-beam distortion, while the curved wall of a traditional spherical cell would act as a thin lens. The edge where the flat window joins to the cell body is intentionally rounded during fabrication. This rounded edge and the tapered exit side both reduce Ne's dipolar field, whose gradient causes spin-relaxation \cite{Cates88}. As shown in Fig.~\ref{fig:NeDipolar}, magnetic field gradient is largest at the edges of a cylinder with uniform longitudinal magnetization pointing along positive \(\hat{z}-\)axis. The magnetic field gradient in the planes of the cylinder's two bases is mainly radial, and thus primarily reduces Ne \(T_1\). Ne’s dipolar field gradient should not have any significant effect other than shortening Ne’s lifetime. One can still use the geometric factor \(\lambda = 8\pi \kappa_{0}/3\) for spherical geometry because the local contact spin interaction between Rb electron spin and Ne nuclear spin is much larger than Ne’s long-range dipolar field, i.e. the enhancement factor \(\kappa_0\) is 35.7 for the former and 1 for the latter. The average longitudinal field within a cylinder deviates from the uniform field inside a uniformly magnetized sphere by at most 3\(\%\), and the actual difference is smaller since the actual cell’s geometry sits somewhere in between a cylinder and a sphere. Optimally the probe beam should horizontally pass through the center of the cell, where Ne's dipolar field is most uniform and transverse field strength is minimal. This also minimizes sensitivity to probe beam translation due to cell dichroism.

\section{\label{sec:level1} Bloch Equation Model}

Like in DC-mode, a set of coupled phenomenological Bloch equations captures the time evolution of electron and nuclear spin polarizations, \(\bf{P^e}\) and \(\bf{P^n}\), for the pulsed comagnetometer:
\begin{align}\label{eqn:Bloch}
\frac{ \partial \textbf{P}^\textbf{e}}{ \partial t} =& \frac{ \gamma _e}{Q(P ^e)}(\textbf{B}+ \lambda M^n_0\textbf{P}^\textbf{n}+\textbf{L}+\boldsymbol\beta^\textbf{e}) \times  \textbf{P}^\textbf{e}+\boldsymbol\Omega \times\textbf{P}^\textbf{e}\nonumber \\&+\frac{1}{Q(P^e)}(R_p\textbf{s}_p+R_m\textbf{s}_m+R^{en}_{se}\textbf{P}^\textbf{n}-R^{e}_{tot}\textbf{P}^\textbf{e})\nonumber
\\\frac{ \partial \textbf{P}^n}{ \partial t} =& \gamma _n(\textbf{B}+ \lambda M^e_0\textbf{P}^\textbf{e}+\boldsymbol\beta^\textbf{n}) \times  \textbf{P}^\textbf{n}+\boldsymbol\Omega \times\textbf{P}^\textbf{n}\nonumber \\&+R^{ne}_{se}\textbf{P}^\textbf{e}-R^{n}_{tot}\textbf{P}^\textbf{n},
\end{align}
where \(\bf{B}\) is the magnetic field, $\boldsymbol\beta^\textbf{e}$ and $\boldsymbol\beta^\textbf{n}$ are the anomalous magnetic-field-like spin interactions of interest coupled to electron spins and nuclear spins respectively, and $\boldsymbol\Omega$ is the inertial rotation rate. \(M^e_0\) and \(M^n_0\) are the magnetizations of electron and nuclear spins at full spin polarization. \(\bf{L}\) is the effective magnetic field affecting only electron spins due to light shift from pump and probe lasers, and \(\bf{s}_p\) and \(\bf{s}_m\) are the directions and magnitudes of their photon spin polarization respectively. Ideally \(|\bf{s}_p|=\)1, and \(|\bf{s}_m|=\)0, corresponding to circular and linear polarization. Relevant gyromagnetic ratios are $\gamma_e$=2$\pi$ 2.8$\times$10$^6$ Hz/G, and $\gamma_n$=2$\pi$ 336.3 Hz/G for $^{21}$Ne. The slowing down factor \(Q(P^e)\), a signature of SERF regime that modifies electron precession frequency, ranges from 6 at \(P^e=0\) to 4 at \(P^e=1\) for alkali-metal isotopes with \(I=3/2\)  \cite{Savukov05}. 

The rates labelled as $R$ with various superscripts and subscripts affect the sensitivity of the comagnetometer and the effectiveness of suppression by noble-gas nuclear spins. \(R_p\) and \(R_m\) are the photon absorption rate for unpolarized alkali atoms of pump and probe light, and depend on the lights' spectral profiles and the transition line shape \cite{walker97}. Within the probe beam's optical path across the cell, both \(R_p\) and \(R_m\) have spatial variations. Such variations should be minimized for optimal performance, as the comagnetometer signal is maximized when all of the atoms coherently precess under the same condition. \(R_p\) also has time dependence, dropping to zero during dark time. \(R^{en}_{se}\) and \(R^{ne}_{se}\) are spin exchange rates representing the  transfer of angular momentum between the two spin species via collisions, during which the noble-gas nuclear spins get polarized. They can be estimated from time-dependent perturbation theory \cite{walker89}, and were experimentally measured for the Rb-Ne pair \cite{Ghosh10}. The total relaxation rate of alkali spins is \(R^e_{tot}=R_p+R_m+R^{en}_{se}+R^{ee}_{se}+R^e_{sd}\), where \(R^{ee}_{se}\) and \(R^e_{sd}\) represent two additional relaxation processes due to spin-exchange collisions between alkali-metal atoms and spin-destruction collisions of alkali-metal atoms with themselves, with noble-gas atoms, and with quenching \(N_2\). The \(R^{ee}_{se}\) term only affects transverse coherence time. Since alkali-metal spins must precess in a finite, i.e. non-zero, longitudinal magnetic field to generate the oscillatory signal of the pulsed comagnetometer, \(R^{ee}_{se}\) is not completely eliminated as in a SERF magnetometer. Nonetheless, its contribution to transverse relaxation remains small as we measured comparable Rb \(T_1\) and \(T_2\) when Ne polarization is kept at zero. This is implemented by rapidly reversing pump laser circular polarization handedness, in a longitudinal magnetic field with equal magnitude as \(B^e\) at the comagnetometer compensation point. The total relaxation rate of noble-gas nuclear spins  \(R^n_{tot}\) has contribution from gas relaxation due to Ne's electric quadrupole moment \cite{Adrian65,Ghosh10}, as well as surface relaxation in the presence of local electric field gradients on the cell wall \cite{Tannoudji63,Wu88}. It also includes contribution from magnetic field gradient due to our cell-geometry.

Numerical simulation with Eqn.~\ref{eqn:Bloch} verified that, akin to the DC-mode comagnetometers, the pulsed comagnetometer exhibits strongly coupled spin dynamics and self-compensating behavior at the compensation point \(\bf{B}\)\(=B^c\hat{z}=-(B^n+B^e)\hat{z}\), where \(B^e\) is determined by the time-averaged electron spin polarization over one pump cycle \(P^e_{eq}\). As shown in Fig.~\ref{fig:RawSignal}(b), at the compensation point, the comagnetometer signal in response to a transverse magnetic field is suppressed by a factor of about 800 compared to response to a transverse anomalous field of the same magnitude. Anomalous fields along \(\hat{x}-\) and  \(\hat{y}-\)axis result in signals that are \(\pi/2\) out of phase,  which means \(A\) and \(B\) in Eqn.~\ref{eqn:fittingForm} can each be made proportional to couplings along one transverse axis with properly chosen \(\phi_0\), thus producing simultaneous dual-axis measurement. For comparison, numerically simulated curves for the DC-mode are also included in Fig.~\ref{fig:RawSignal}(b). All parameters in the DC-mode model are the same as those in the pulsed mode model, except for \(R_p\), which is set to the value that makes the DC-mode equilibrium \(P^e_z\) equal to the pulsed mode \(P^e_{eq}\). The DC-mode is insensitive to \(B_y\) and anomalous spin interactions along the \(\hat{x}\)-axis, and \(B_x\) is suppressed by a factor of about 3900 compared to \(\beta_y\).

\begin{figure}
\includegraphics[width=8.5cm, keepaspectratio]{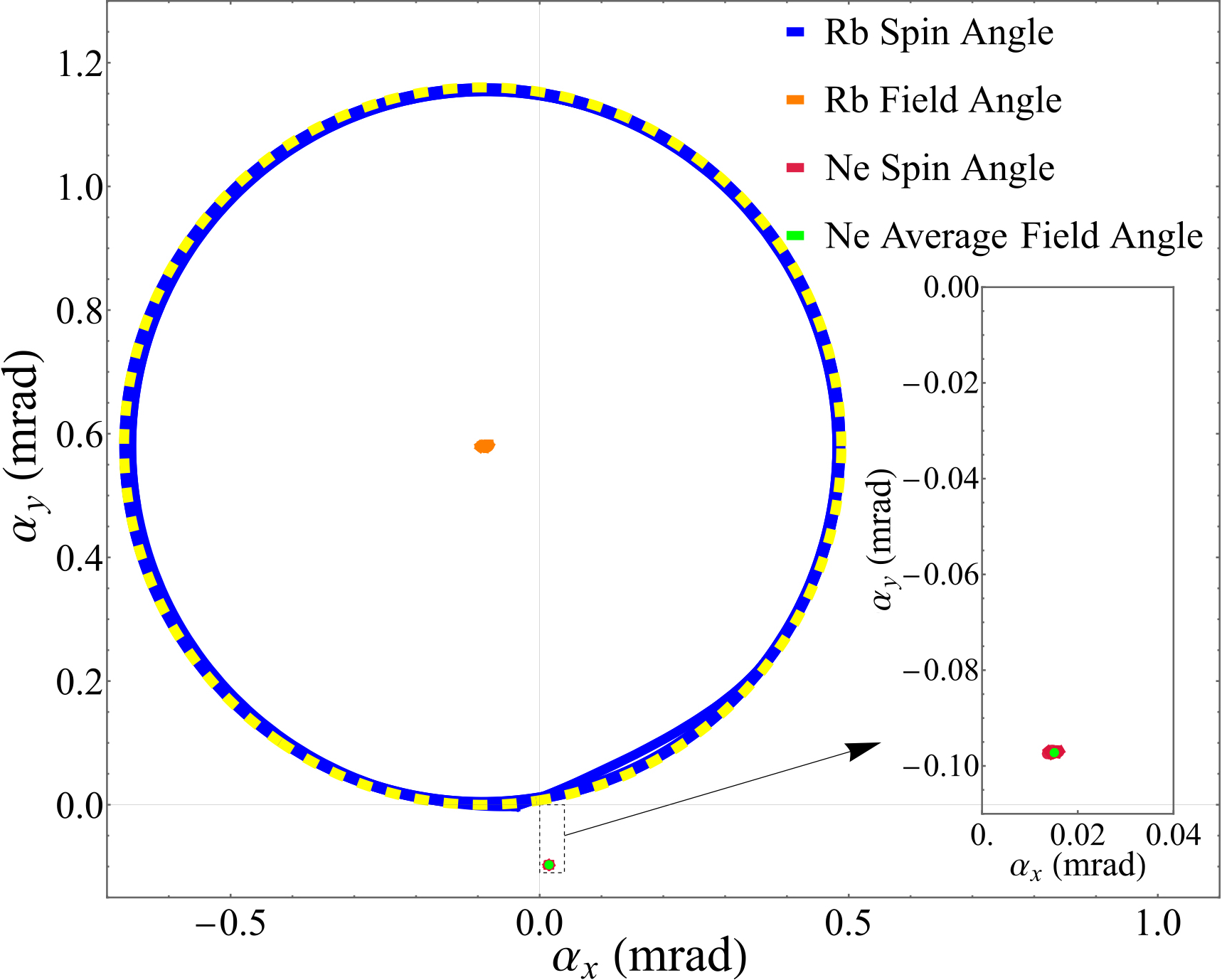}
\caption{\label{fig:LineUp} Numerically simulated spin trajectory at the compensation point in response to \(\beta^n_{y}=10^{-7}\) G over one pump period, and the associated effective fields felt by the two spin species, viewed in the transverse plane. All are expressed as polar angles as defined in Eqn.~\ref{eqn:NeAngleNeBeff}-\ref{eqn:RbMeanAngleRbBeff} and the text following them. The yellow dashed line is a circle centered at the time-averaged Rb field angle. Rb spin angle trajectory largely overlaps with this circle during dark time, and deviates from it during the pumping pulse. Ne spin angle largely aligns with its average field angle. Without the probe beam the system has rotational symmetry around the longitudinal axis. Spin trajectories in response to \(\beta^n_{x}=10^{-7}\) G correspond to the above trajectories rotated by 90 degrees clockwise around the origin. }
\end{figure}

To understand qualitatively some of the essential features for pulsed operation we develop an approximate analytical model. Under typical operating conditions spin couplings in the transverse plane are small, so they barely affect longitudinal polarization, i.e.
\begin{equation}
    \bar{P^e_{z}}\simeq P^e_{eq} =\frac{B^e}{\lambda M^e_0}, \;\text{and }{P^n_{z}}\simeq P^n_{eq}=\frac{B^n}{\lambda M^n_0}.
\end{equation}
 \(\bar{P^e_{z}}\) is the longitudinal component of the time-averaged Rb polarization given by 
\begin{equation}
\bar{P^e_{\perp/z}}=\frac{\int_0^{T_{cycle}} P^e_{\perp/z} dt}{T_{cycle}},
\end{equation}
where \(T_{cycle}=T_{pump}+T_{dark}\) is the period of one pump cycle, and the complex term \(P_{\perp}=P_x+iP_y\) is introduced for simplified notation in the subsequent calculations. The transverse spin polarization can be expressed in terms of polar angles. Numerical simulations shown in Fig.~\ref{fig:LineUp} indicate that at the compensation point after reaching equilibrium the Ne polarization angle is aligned with the angle of the effective field it experiences due to Rb magnetization and external fields that couple to Ne, while Rb polarization precesses around its effective field angle due to Ne magnetization and external fields that couple to Rb. This observation leads to the following equations
\begin{align}\label{eqn:NeAngleNeBeff}
\frac{P^n_{\perp}}{P^n_z}&=\frac{\lambda M^e_0 \bar{P^e_{\perp}}+B_{\perp}+\beta^n_{\perp}}{B^c+\lambda M^e_0\bar{P^e_{z}}}\equiv \alpha^n_\perp,\\\label{eqn:RbMeanAngleRbBeff}
\frac{\int_{0}^{T_{Rb}}\frac{P^e_{\perp}}{P^e_z} dt}{T_{Rb}}&=\frac{\lambda M^n_0 P^n_{\perp}+B_{\perp}+\beta^e_{\perp}}{B^c+\lambda M^n_0 P^n_{z}}\equiv \alpha^e_\perp,
\end{align}
where \(t=0\) corresponds to the end of the pumping pulse, \(T_{Rb}\) is the time taken for Rb polarization to travel around the circle once, and components of Ne polarization are treated as constants since they have little time dependence. The field angle \(\alpha^e_\perp\) is defined as the ratio of the transverse components of effective field felt by Rb, \(B^{Rb-eff}_\perp\), and its corresponding longitudinal component, \(B^{Rb-eff}_z\). The field angle \(\alpha^n_\perp\) is similarly defined. The time-dependent spin angles as shown in Fig.~\ref{fig:LineUp} are \([\alpha_x,\alpha_y]=[P^{e/n}_x(t),P^{e/n}_y(t)]/P^{e/n}_z(t)\). If one can express \(\bar{P^e_{\perp}}\) in terms of \(\alpha^e_\perp\), \(P^n_{\perp}\) can be solved for by plugging Eqn.~\ref{eqn:RbMeanAngleRbBeff} into Eqn.~\ref{eqn:NeAngleNeBeff}. With a known \(P^n_{\perp}\), Rb field angle is also known. Radius of the Rb spin angle circle is thus equal to the angular distance between Rb field angle and the Rb spin angle at the end of pumping pulse. Assuming \(R_p\) is much larger than the other rates, at the end of the strong pumping pulse \(P^e_{\perp}=0\) and \(P^e_z=1\). Then the distance between Rb field angle and the origin is equal to the initial oscillation amplitude of \(P^e_x\) as shown in Fig.~\ref{fig:RawSignal}(b).

In order to arrive at an analytic expression connecting \(\bar{P^e_{\perp}}\) and \(\alpha^e_\perp\), one makes the following assumptions. First, the pumping pulse is extremely strong and short, \(R_p\rightarrow\infty\), \(T_{pump}\rightarrow 0\), and \(T_{cycle}\rightarrow T_{dark}\). Second, after reaching \(P^e_{\perp}=0\) and \(P^e_z=1\) at the end of the pumping pulse, during \(T_{dark}\) \(P^e_z\) exponentially decays following \(P^e_z(t)=e^{-(R_{dark}/Q)t}\), where \(R_{dark}=R^e_{tot}-R_p-R^{ee}_{se}\). The average longitudinal polarization is thus \( \bar{P^e_{z}}=P^e_{eq}=T_1(1-e^{-T_{dark}/T_1})/T_{dark}\). Third, Rb atoms are in the near-SERF regime, so \(R^{ee}_{se}\simeq 0\) and transverse relaxation time \(T_2\) equals \(T_1\). Fourth, the slowing down factor \(Q\) is a constant, since it has a small range from 4 to 6. Hence the longitudinal relaxation time can be expressed as \(T_1=Q/R_{dark}\). The transverse polarization during dark time can then be calculated from the Bloch equation
\begin{equation}\label{eqn:PeBloch}
    {P^e_{\perp}}'=\frac{1}{Q}[(i\gamma_eB^{Rb-eff}_z-R_{dark})P^e_{\perp}-i\gamma_eB^{Rb-eff}_{\perp}e^{-\frac{R_{dark}}{Q}t}]
\end{equation}
with initial condition \(P^e_{\perp}(0)=0\). The solution is
\begin{equation}\label{eqn:PeBlochSolution}
    P^e_{\perp}(t)=\frac{B^{Rb-eff}_{\perp}}{B^{Rb-eff}_z}e^{-\frac{t}{T_1}}(1-e^{i\omega_{Rb}t}),
\end{equation}
where \(\omega_{Rb}=\gamma_e B^{Rb-eff}_z/Q\). The field angle \(\alpha^e_{\perp}\) is simply
\begin{equation}
    \frac{\int_{0}^{T_{Rb}}\frac{P^e_{\perp}}{P^e_z} dt}{T_{Rb}}= \frac{\int_{0}^{2\pi /\omega_{Rb}}\frac{B^{Rb-eff}_{\perp}}{B^{Rb-eff}_z}(1-e^{i\omega_{Rb}t}) dt}{2\pi /\omega_{Rb}}=\frac{B^{Rb-eff}_{\perp}}{B^{Rb-eff}_z}.
\end{equation}
For \(\bar{P^e_{\perp}}\) one first considers the simpler case where during the dark time Rb polarization traverses its circular spin angle trajectory an integer number of times, \(T_{dark}=n T_{Rb} = 2\pi n/\omega_{Rb}\). After integration the sought-after expression is 
\begin{align}\label{eqn:PeAlphae}
    \bar{P^e_{\perp}}&=\frac{B^{Rb-eff}_{\perp}}{B^{Rb-eff}_z}\frac{(1-e^{-\frac{2n\pi}{T_1\omega_{Rb}}})(T_1\omega_{Rb})^2}{2n\pi(i+T_1\omega_{Rb})}\\
    &= \alpha^e_\perp P^e_{eq}\frac{T_1\omega_{Rb}}{i+T_1\omega_{Rb}}.\nonumber
\end{align}

Using the expression above to evaluate Eqn.~\ref{eqn:NeAngleNeBeff} and Eqn.~\ref{eqn:RbMeanAngleRbBeff}, transverse components of Ne polarization are given by
\begin{align}
P^n_x&=\frac{-B_x-\beta^n_{x}+(\beta^e_{y}-\beta^n_{y})T_1\omega_{Rb}}{\lambda M^n_0},\nonumber\\
P^n_y&=\frac{-B_y-\beta^n_{y}+(-\beta^e_{x}+\beta^n_{x})T_1\omega_{Rb}}{\lambda M^n_0}.
\end{align}
Rb field angles are then given by
\begin{align}\label{eqn:alphaRbPerp}
\alpha_x^e=&\frac{\lambda M^n_0 P^n_x+B_x+\beta^e_{x}}{B^c+\lambda M^n_0 P^n_{eq}}=\frac{\beta^e_x-\beta^n_{x}+(\beta^e_{y}-\beta^n_{y})T_1\omega_{Rb}}{-\lambda M^e_0 P^e_{eq}},\nonumber\\
&\alpha_y^e=\frac{\beta^e_y-\beta^n_{y}+(-\beta^e_{x}+\beta^n_{x})T_1\omega_{Rb}}{-\lambda M^e_0 P^e_{eq}}.
\end{align}
Again, for high \(R_p\) the quadrature sum of \(\alpha_x^e\) and \(\alpha_y^e\) is the initial oscillation amplitude of the pulsed comagnetometer signal. From Eqn.~\ref{eqn:alphaRbPerp} it is obvious that the pulsed comagnetometer is insensitive to transverse magnetic field \(B_{\perp}\), thanks to compensation from Ne polarization. Sensitivity to anomalous fields \(\beta^n_{\perp}\) and \(\beta^e_{\perp}\) is the same with opposite sign. This sensitivity is given by
\begin{equation}\label{eqn:sensitivity}
   \frac{\sqrt{1+(T_1\omega_{Rb})^2}}{\lambda M^e_0 P^e_{eq}}\simeq \frac{T_1\omega_{Rb}}{\lambda M^e_0 P^e_{eq}}=\frac{\frac{Q}{R_{dark}}\frac{\gamma_eB^e}{Q}}{B^e}=\frac{\gamma_e}{R_{dark}}.
\end{equation}
For comparison, the DC-mode comagnetometer's sensitivity to \(\beta^{e/n}_y\) is \(\gamma_e P^e_0/R^e_{tot}\) at the compensation point, which is maximized at equilibrium polarization \(P^e_0=0.5\). Under similar operating conditions the pulsed comagnetometer's initial signal response to an anomalous field is four times that of the DC-comagnetometer, whose \(R^e_{tot}\) has contribution from \(R_p\), which is equal to the sum of all other relaxation rates to maintain \(P^e_0=0.5\).

\begin{figure}
\includegraphics[width=8cm, keepaspectratio]{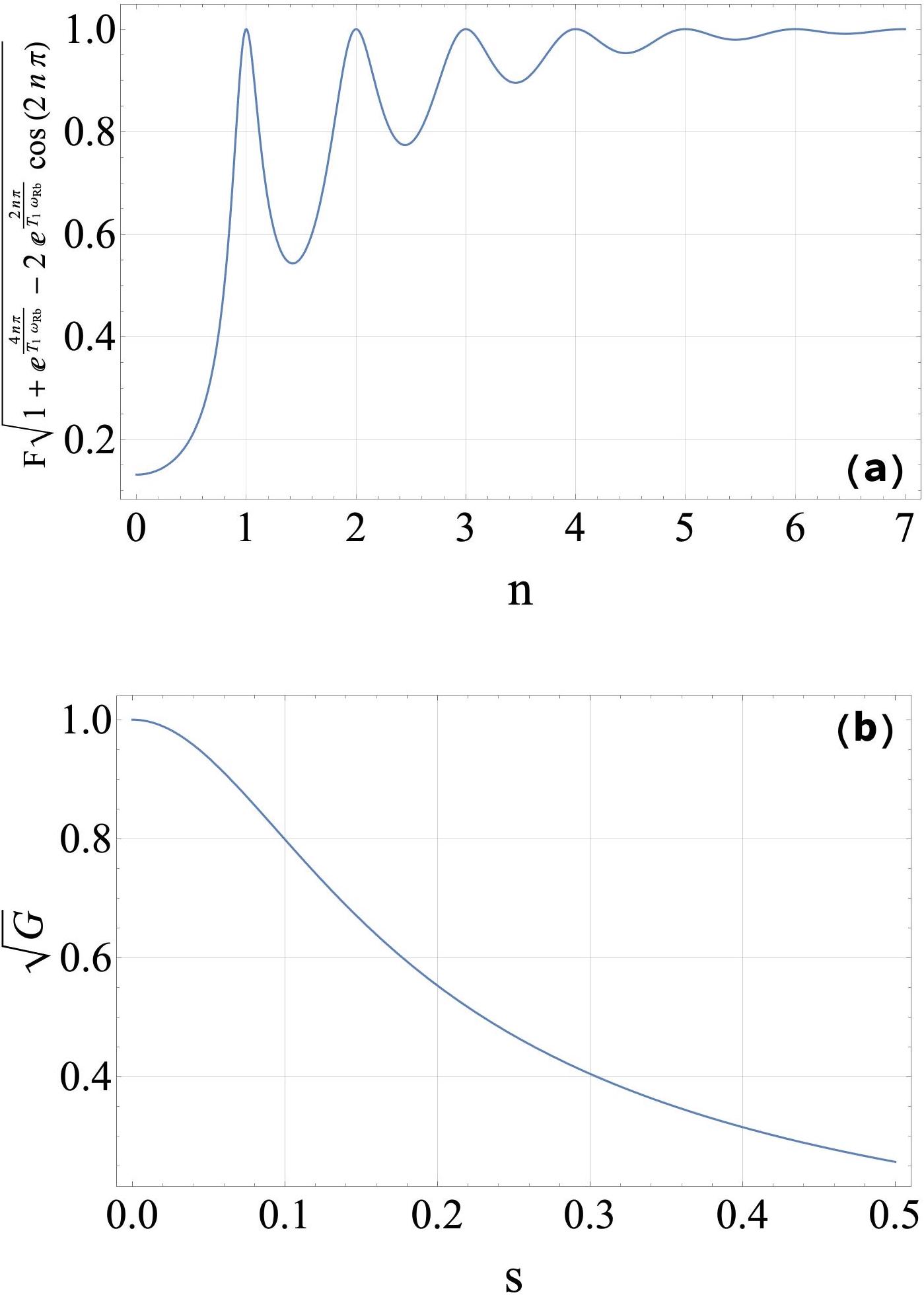}
\caption{\label{fig:SensFactor} Analytical pulsed comagnetometer sensitivity in unit of \(\gamma_e/R_{dark}\) for typical experimental values \(T_1=6\) ms, \(\omega_{Rb}=2\pi 200\) Hz, and \(P^e_{eq-dark}=0.5\). (a) Pulsed comagnetometer sensitivity as a function of number of oscillation periods \(n\). (b) Pulsed comagnetometer sensitivity as a function of \(s\). Smaller \(s\) corresponds to shorter pumping pulse.}
\end{figure}

Due to drifts of experimental conditions it is impractical to enforce \(n\) to be an integer. Without this assumption the Rb field angles are given by
\begin{align}\label{eqn:alphaRbNonInteger}
    \alpha_x^e=&\frac{F}{-\lambda M^e_0P^e_{eq}} \left[B_1\left(e^{\frac{2n\pi}{T_1\omega_{Rb}}}-\cos{2n\pi}\right)-B_2\sin{2n\pi}\right],\nonumber\\
     \alpha_y^e=&\frac{F}{-\lambda M^e_0P^e_{eq}}  \left[B_2\left(e^{\frac{2n\pi}{T_1\omega_{Rb}}}-\cos{2n\pi}\right)+B_1\sin{2n\pi}\right],
\end{align}
where for simplified notation the following substitutions are defined
\begin{align}
    F =&\frac{e^{\frac{2n\pi}{T_1\omega_{Rb}}}-1}{1+e^{\frac{4n\pi}{T_1\omega_{Rb}}}-2e^{\frac{2n\pi}{T_1\omega_{Rb}}}\cos{2n\pi}},\nonumber\\
    B_1=&\beta^e_x-\beta^n_x+(\beta^e_y-\beta^n_y)T_1\omega_{Rb},\nonumber\\
    B_2=&\beta^e_y-\beta^n_y+(-\beta^e_x+\beta^n_x)T_1\omega_{Rb}.
\end{align}
It is still true that the pulsed comagnetometer signal is insensitive to transverse magnetic fields. Sensitivity to anomalous fields is modified to 
\begin{equation}\label{eqn:sensitivityNonInteger}
    \frac{\sqrt{1+(T_1\omega_{Rb})^2}}{\lambda M^e_0 P^e_{eq}} F \sqrt{ 1+e^{\frac{4n\pi}{T_1\omega_{Rb}}}-2e^{\frac{2n\pi}{T_1\omega_{Rb}}}\cos{2n\pi}}.
\end{equation}
When \(n\) is an integer Eqn.~\ref{eqn:alphaRbNonInteger} reduces to Eqn.~\ref{eqn:alphaRbPerp} and Eqn.~\ref{eqn:sensitivityNonInteger} reduces to Eqn.~\ref{eqn:sensitivity}, as can be seen in Fig.~\ref{fig:SensFactor}(a). Not having an integer number of periods during dark time reduces sensitivity.

In practice the pumping pulse cannot be infinitely short. Still assuming \(R_p\rightarrow\infty\) and \(n\) is an integer, but \(T_{pump}\neq 0\), \(\bar{P^e_{\perp}}\) should now equal its dark-time-only value multiplied by \(T_{dark}/T_{cycle}\), since during the pulse transverse polarization is zero. \(\alpha^e_\perp\) remains unchanged since the integration in Eqn.~\ref{eqn:RbMeanAngleRbBeff} excludes pumping time. \(P^e_{eq}\) changes to \((P^e_{eq-dark}T_{dark}+T_{pump})/T_{cycle}\). Letting \(P^e_{eq-dark}T_{dark}/T_{cycle} = (1-s)P^e_{eq}\), Eqn.~\ref{eqn:PeAlphae} is modified to 
\begin{equation}
\bar{P^e_{\perp}}=(1-s)\alpha^e_\perp P^e_{eq}\frac{T_1\omega_{Rb}}{i+T_1\omega_{Rb}},
\end{equation} 
and Rb field angles are given by
\begin{align}
    \alpha^e_x=&\frac{G}{-\lambda M^e_0 P^e_{eq}}[(\beta^e_y-\beta^n_y)(1-s)T_1\omega_{Rb}\\
    &+(\beta^e_x-\beta^n_x)(1+s {T_1}^2{\omega_{Rb}}^2)],\nonumber\\
     \alpha^e_y=&\frac{G}{-\lambda M^e_0 P^e_{eq}}[(\beta^e_y-\beta^n_y)(1+s {T_1}^2{\omega_{Rb}}^2)\nonumber\\
     &-(\beta^e_x-\beta^n_x)(1-s)T_1\omega_{Rb}],\nonumber
\end{align}
where \(G\) is defined as
\begin{equation}
    G=\frac{1}{1+s^2 {T_1}^2{\omega_{Rb}}^2}.
\end{equation}
Again, transverse magnetic fields are suppressed. The sensitivity becomes
\begin{align}\label{eqn:sensitivityLongPulse}
   \frac{\sqrt{1+(T_1\omega_{Rb})^2}}{\lambda M^e_0 P^e_{eq}}\sqrt{G}.
\end{align}
An explicit expression for \(s\) is 
\begin{equation}
    s=\frac{T_{pump}}{T_{pump}+T_1(1-e^{-T_{dark}/{T_1}})}.
\end{equation}
As shown in Fig.~\ref{fig:SensFactor}(b), Eqn.~\ref{eqn:sensitivityLongPulse} reduces to Eqn.~\ref{eqn:sensitivity} for extremely short pumping pulse where \(s=0\). Increasing pumping pulse length decreases sensitivity. Under typical experimental conditions \(s=0.38\), for which sensitivity is reduced by a multiplicative factor of 0.33.

To analyze comagnetometer response to a tilted pump beam, it is useful to define a \(uvw-\)coordinate system whose \(\hat{w}-\)axis is parallel to the pump beam's direction, and whose origin is the same as that of the \(xyz-\)coordinate system. Let the unit vector \(\bf{\hat{s}_p}\) denoting the pump's direction have \(xyz-\)coordinates \((\text{sin}\theta \text{cos}\phi, \text{sin}\theta \text{sin}\phi,\text{cos}\theta)\), where \(\theta\) is the angle down from \(\hat{z}-\)axis, and \(\phi\) is the azimuthal angle from \(\hat{x}-\)axis. The \(uvw-\)coordinate system can be obtained from a rotation by angle \(\theta\) around the axis \(\bf{\hat{z}\times\hat{s}_p}/\)\(\|\bf{\hat{z}\times\hat{s}_p}\|\)\(=(-\text{sin}\phi,\text{cos}\phi,0)\). For a nominally aligned configuration, \(B_x=B_y=0\) and \(B_z=B^c\). Then in \(uvw-\)coordinates the spins experience effective field \((B_u,B_v,B_w)=(-B^c\text{sin}\theta\text{cos}\phi,-B^c\text{sin}\theta\text{sin}\phi, B^c\text{cos}\theta)\). For a tilted pump beam with \(\theta \neq 0\), the longitudinal field in \(uvw-\)coordinates deviates from the compensation point by \((1-\text{cos}\theta)B^c\), and the spins feel non-zero transverse fields. Hence it is of interest to analyze the pulsed comagnetometer's suppression of transverse magnetic fields for \(B_w\) detuned from \(B^c\) by \(\delta B_w\). Adopting the same set of assumptions used to arrive at Eqn.~\ref{eqn:sensitivity}, and expanding to lowest order in \(\delta B_w\), Rb field angles are given by 
\begin{equation}
    (\alpha_u^e,\alpha_v^e) = \left(\frac{B_u+B_vT_1\omega_{Rb}}{B^e}\frac{\delta B_w}{B^n},\frac{B_v-B_uT_1\omega_{Rb}}{B^e}\frac{\delta B_w}{B^n}\right).
\end{equation} 
A pump beam tilt is suppressed in the same way as a magnetic field tilt. Plugging in the effective fields, signal oscillation amplitude in response to tilt \(\theta\) is given by Eqn.~\ref{eqn:sensitivity} multiplied by
\begin{equation}
    (1-\cos\theta)\sin\theta\frac{{B^c}^2}{B^n}\simeq \frac{\theta^3{B^c}^2}{2 B^n}+\mathcal{O}(\theta^5).
\end{equation}
Depending on \(\phi\), the projection of this amplitude along the probe beam, i.e. the \(\hat{x}-\)axis, is modified by a factor ranging from \(\cos{\theta}\) to 1. The DC-mode comagnetometer also suppresses \(B_{u/v}\) in the \(uvw-\)frame. However, its DC signal still responds to pump beam tilts, because projecting the \(\hat{w}-\)component of electron spin polarization along the probe beam results in a DC-offset \(P^e_0\sin\theta\sin\phi\).  While \(\sin\phi=0\) for pump beam tilts into \(\hat{y}-\)axis, beam pointing fluctuations are not constrained to lie along any particular axis, and thus contribute to \(1/f\) noise in DC-mode. In the pulsed comagnetometer this projection appears as an exponential decay without affecting oscillation amplitudes. To realize the above equivalence of suppression of pump laser tilt to suppression of magnetic field tilt, a well-collimated pump beam is of critical importance to the pulsed comagnetometer, since otherwise the \(\hat{w}-\)axis is ill-defined, and so are the equivalent fields. The entrance window of our cell requires a pump beam of 0.01 rad radial divergence to become collimated after passing through it. Since the window's effective focal length is different along different directions, the radial divergence within the cell could fluctuate by about 2 mrad. As shown in Section IV, this is large enough to be a limiting factor on the comagnetometer's long-term sensitivity.

\section{\label{sec:level1}Operation}
An elaborate zeroing procedure based on quasi-steady-state modulation of magnetic fields and of optical pumping rate and wavelength was developed for the initial K-\(^3\)He comagnetometer \cite{kornackThesis} to minimize the steady-state signal's sensitivity to magnetic fields, light shifts, and alignment imperfections. For the pulsed comagnetometer, it suffices to zero a smaller subset of the above, namely \(\bf{B}\) and light shift from the probe laser, due to the pulsed scheme's built-in insensitivity to other systematics. However, the quasi-steady-state modulations used in the DC-mode zeroing procedure generally do not transfer to the pulsed case due to more complicated signal dependence on various terms as a result of having dual-axis sensitivity. To set longitudinal magnetic field to \(B^c\), where suppression of magnetic fields is at its maximum, we apply a slow sinusoidal modulation of \(B_x\) or \(B_y\) at a frequency \(\omega \ll \gamma_n B^n\). Very slow modulation of the transverse fields can be thought of as an active rotation of the magnetometer
axes resulting in the tilt of the \(B_z\) field into the transverse axis, while the tilt of the lasers is suppressed at the compensation point. Therefore, it also provides a robust calibration for inertial rotation or anomalous fields \cite{brownThesis}. Additionally, this modulation allows us to find the proper \(\phi_0\) in Eqn.~\ref{eqn:fittingForm} that assigns all \(\hat{x}-\)response to \(A\) and \(\hat{y}-\)response to \(B\), or vice versa. During long-term measurement we apply such modulation every few hundred seconds to compensate for \(B^c\) drifts over time, similar to in DC-mode.

\begin{figure}
\includegraphics[width=8cm, keepaspectratio]{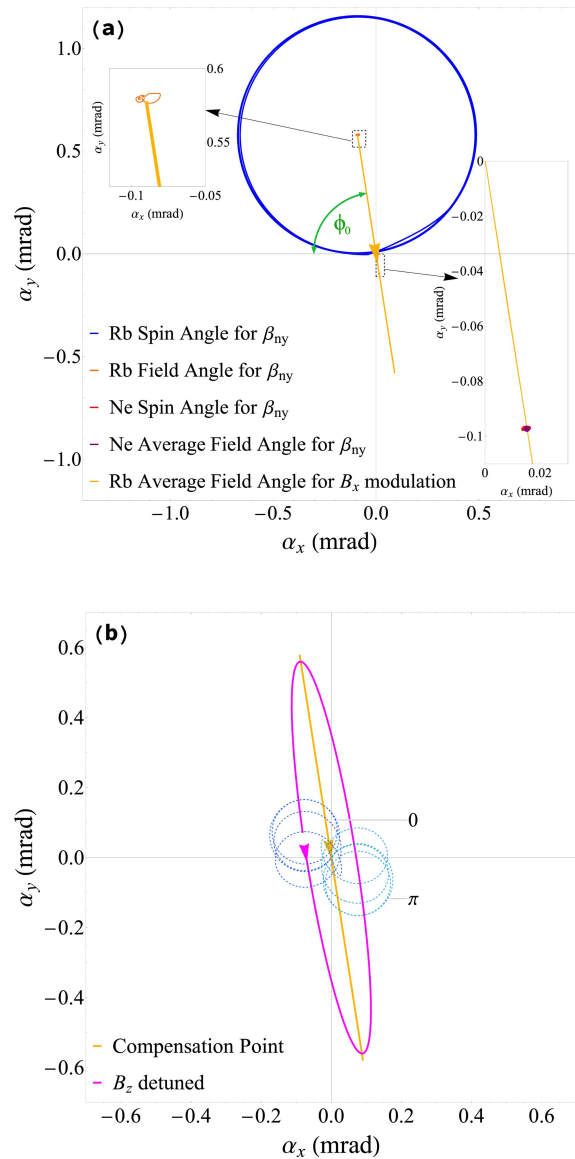}
\caption{\label{fig:slowBx} Numerically simulated effective fields felt by Rb spins with slow modulation \(B_x = B_{x0}\cos{\omega t}\), where \(B_{x0} = (\gamma_n B^n/\omega)10^{-7}\) G and \(\omega = 2\pi 0.01\) rad/s. Arrows are placed on field angle lines to indicate the angle at \(t=0\) (arrowhead bottom), and where it moves as \(t\) increases (arrowhead tip). (a) At the compensation point. Trajectories in response to \(\beta^n_{y}=10^{-7}\) G, already shown in Fig.~\ref{fig:LineUp}, are included for comparison. (b) \(B_z\) detuned from the compensation point by a few \(\mu\)Gs. Rb average field angle at the compensation point is included for comparison. Blue dotted lines represent the time-dependent Rb spin angles during pump cycles starting at different \(t\). The label values for these lines are equal to \(\omega t\).}
\end{figure}

Numerical simulation using the Bloch equations shown in Fig.~\ref{fig:slowBx} illustrates the evolution of the spins during a slow \(B_x\) modulation, compared with that in response to an anomalous field signal \(\beta^n_{y}\). When the comagnetometer is at the compensation point, during the slow \(B_x\) modulation the oscillation amplitude of Rb signal is mostly proportional to the derivative of the \(B_x\) field. The effective field felt by Rb spins follows a straight line in the \(\{\alpha_x,\alpha_y\}\) plane, crossing the origin. This means for all pump cycles during the modulation, the transient comagnetometer signal has the same phase. Thus it is possible to find \(\phi_0\) such that only one of \(A\) and \(B\) in Eqn.~\ref{eqn:fittingForm} responds to the modulation. The direction of the Rb effective field angle line indicates that this phase angle cleanly separates \(\hat{x}\) and \(\hat{y}\) response of the comagnetometer, since it is aligned with the response to an anomalous field \(\beta^n_{y}\). In the ideal scenario where the pumping pulse pins the spins to the \(\hat{z}-\)axis and the fitting starts right at the end of the pulse, where \(t\) is set to zero, the angle \(\phi_0\) as drawn in Fig.~\ref{fig:slowBx}(a) is the fitting phase making \(A\) sensitive to \(\hat{x}-\)couplings and \(B\) sensitive to \(\hat{y}-\)couplings. For a modulation amplitude of \(\beta^n_{y}\gamma_n B^n/\omega\), one maximum angle on the Rb average field angle line coincides with the time-averaged Rb field angle for \(\beta^n_{y}\), as shown in the upper left inset of Fig.~\ref{fig:slowBx}(a). Since Rb spins precess around this field angle, the maximum signal amplitude during slow \(B_x\) modulation is the same as signal amplitude in response to a \(\beta^n_y\) whose magnitude is \(\omega/\gamma_nB^n\) that of the applied \(B_x\) amplitude. This allows comagnetometer calibration using slow \(B_x\) modulation. When the \(B_z\) field is not at the compensation point, the spin angle follows an elliptical shape, so a fixed \(\phi_0\) cannot eliminate either \(A\) or \(B\) throughout the modulation, as shown in Fig.~\ref{fig:slowBx}(b).

\begin{figure}
\includegraphics[width=8.5cm, keepaspectratio]{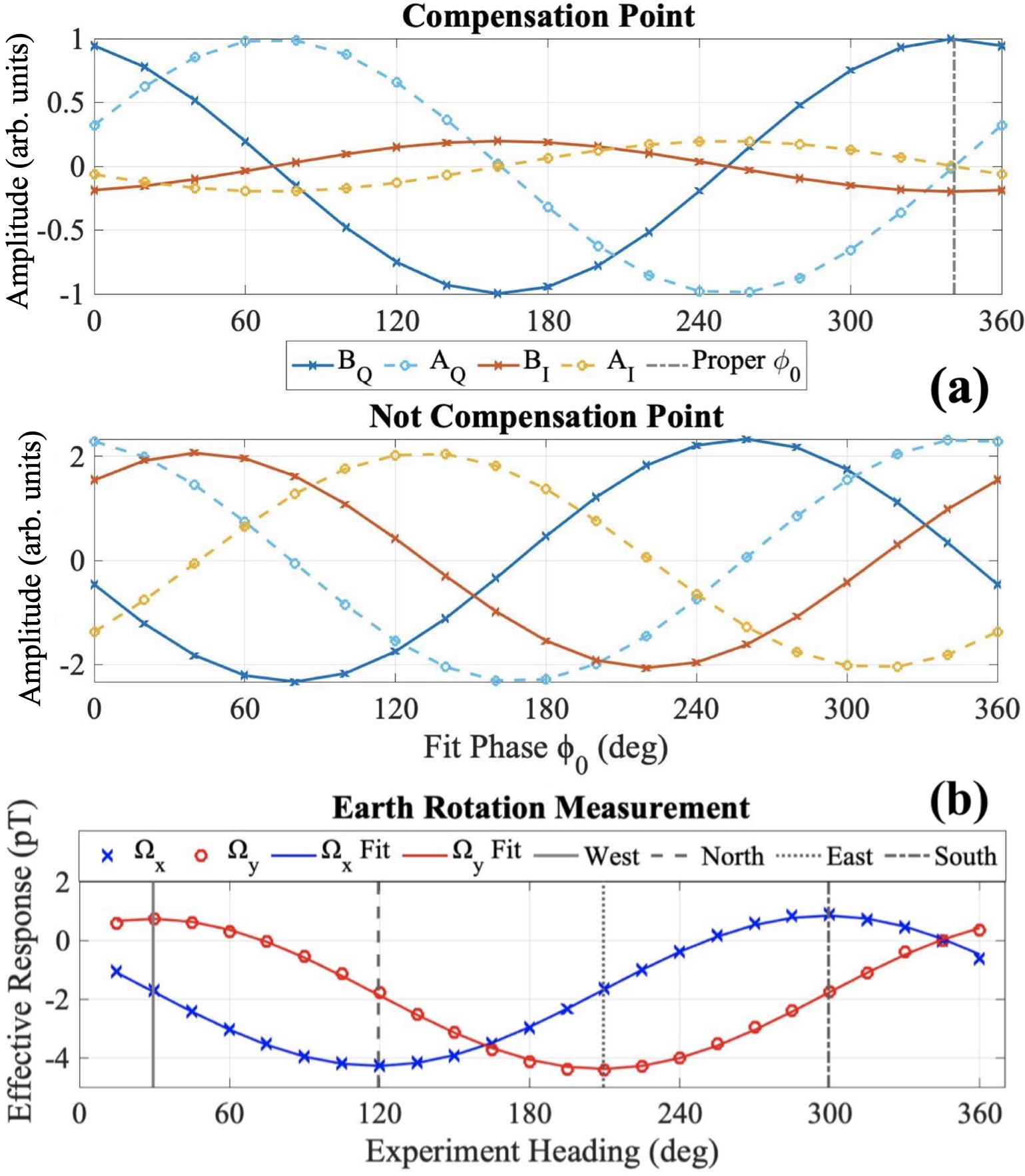}
\caption{\label{fig:CPID} (a) Experimental amplitudes of in-phase and quadrature components of \(A(t)\) and \(B(t)\) in response to \(B_x(t)=B_{x0}\text{cos}(\omega t)\). Top panel: At the compensation point, the leading order response is mostly in quadrature with \(B_x\) modulation. At the proper \(\phi_0\), \(A_I=A_Q=0\), which means \(B\) is sensitive to anomalous \(\hat{y}\)-couplings and \(A\) to \(\hat{x}\)-couplings. We impose that \(B_Q>0\) at this phase to get rid of the degeneracy. Bottom panel: Off the compensation point, in-phase and quadrature components as functions of \(\phi_0\) have a relative phase shift, and a proper \(\phi_0\) that separates \(\hat{x}\)- and \(\hat{y}\)-couplings into \(A\) and \(B\) does not exist. (b) Angular velocity from Earth's rotation measured with the pulsed comagnetometer, calibrated to effective magnetic units using calibration from slow \(B_x\) modulation, i.e. \(\Omega_x = A\times\)Calibration and \(\Omega_y = B\times\)Calibration. Error bars are comparable to point size and therefore omitted. The ``\(\Omega_{x/y}\) Fit'' curves are results from fitting the data points to \(\Omega_{\oplus}\sin(\text{Heading}+\theta)+\text{Offset}\), where \(\Omega_{\oplus x}=2.56\) pT, \(\Omega_{\oplus y}=2.56\) pT, \(\theta_x= 208.8\)\textdegree, \(\theta_y= 298.7\)\textdegree, Offset\(_x=-1.72\) pT, and Offset\(_y=-1.83\) pT. Theoretical amplitude \(\Omega_{\oplus}\) is 2.63 pT at the location of the comagnetometer.}
\end{figure}

Experimental data illustrating how slow \(B_x\) modulation help locate the compensation point are shown in Fig.~\ref{fig:CPID}. As shown in Fig.~\ref{fig:slowBx}(a), the bottom of the orange arrowhead, which corresponds to the field angle at \(B_x(t=0)\), is close to the origin but not exactly at the origin. This indicates that the amplitude of transient signal oscillates at the modulation frequency \(\omega\) mostly out-of-phase with respect to the applied \(B_x(t)\), with a small in-phase component, which is also present in DC-mode~\cite{brownThesis}. For the purpose of \(B_z\) detuning zeroing we choose an arbitrary fixed phase \(\phi_0\), use it to fit the transient signal for \(A\) and \(B\), and compute the coefficients of in-phase cos\((\omega t)\) components and quadrature sin\((\omega t)\) components, which we label as \(A_I, A_Q, B_I,\) and \(B_Q\). \(A_I\) is computed from \(A(t)\) as the integral \(\omega/\pi\)\(\int_{0}^{2\pi/\omega} A(t)\cos{\omega t}\,dt\) and \(A_Q\) as \(\omega/\pi\)\(\int_{0}^{2\pi/\omega} A(t)\sin{\omega t}\,dt\). \(B_{I/Q}\) components are similarly defined. By varying the chosen phase the above four components each becomes a function of \(\phi_0\), which we plot in Fig.~\ref{fig:CPID}(a). As expected, when at the compensation point there exists a \(\phi_0\) that renders \(A\) insensitive to the modulation. A deviation of \(B_z\) from the compensation point results in a relative phase shift between \(A_I(\phi_0)\) and \(A_Q(\phi_0)\), so no value of \(\phi_0\) can eliminate \(A=\sqrt{{A_I}^2+{A_Q}^2}\). At the compensation point using the proper \(\phi_0\), \(B(t)\)'s amplitude is connected to the comagnetometer response to anomalous fields as 
\begin{equation}
   A(\beta^n_x) \simeq B_Q \frac{\omega B_{x0}}{\gamma_n B^n}\beta^n_x,\text{ and } B(\beta^n_y) \simeq B_Q \frac{\omega B_{x0}}{\gamma_n B^n}\beta^n_y,
\end{equation}
allowing the calibration of the comagnetometer signal in terms of the known \(\omega\) and \(B_{x0}\). Response to \(\beta^e_{x/y}\) is the same except for a sign flip. Experimentally \(B^n\) is calculated as \(B^c+B^e\), where \(B^e\) can be estimated by extrapolating the curve of transient frequency \(f\) in Eqn.~\ref{eqn:fittingForm} as a function of \(B_z\).

As a check on the pulsed comagnetometer's dual-axis performance and on its calibration obtained from slow \(B_x\) modulation, we compare its signal response to Earth's rotation with theoretical values. By rotating the experiment around its vertical axis, projection of Earth's rotation \(\bf{\Omega}_\oplus\) onto the two sensitive axes oscillates, as shown in Fig.~\ref{fig:CPID}(b),  and dual-axis performance can be observed. Measured amplitudes agree with theoretical ones within the calibration accuracy of 3\(\%\), comparable to that of the DC-mode comagnetometer gyroscope \cite{kornack05}. Same as in DC-mode \cite{brownThesis}, this accuracy is limited by the uncertainty of \(B^n\) due to residual longitudinal magnetic fields inside the magnetic shields. There are offsets in the measured \(\Omega_x\) and \(\Omega_y\), suggesting the presence of non-zeroed transverse magnetic field, since other anomalous couplings are typically orders of magnitude  smaller. However, these offsets, on the order of Earth's rotation rate, are much smaller than typical
DC signals observed with a DC-mode comagnetometer, where no reliable way of finding absolute zero signal exists. The orthogonality of \(\Omega_x\) and \(\Omega_y\) 
is typically much better than 1 \(\degree\).

\begin{figure}
\includegraphics[width=8.5cm, keepaspectratio]{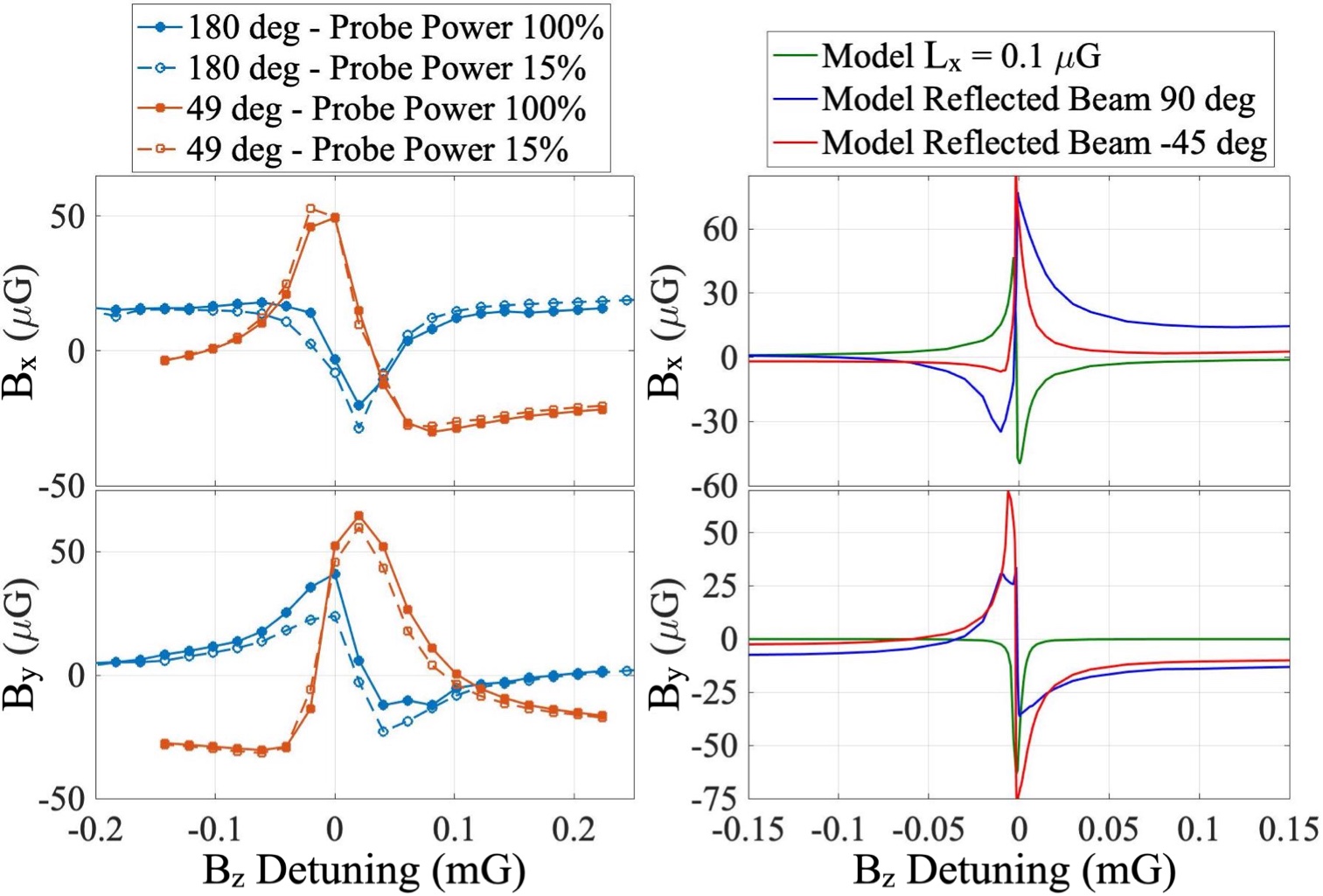}
\caption{\label{fig:dispersion} \(B_x\) and \(B_y\) that zeroes transient signal amplitude as a function of \(B_z\), close to compensation point. Left: Experimental curves showing effect from a rotation of the cell around \(\hat{z}\)-axis with respect to other parts of the experiment. Comparison between high and low probe power shows that probe beam's effect is relatively small. Right: Model curves for \(B_{x/y}\) that zeroes \(L_x\)=0.1 \(\mu\)G, and for different cell orientation, assuming 25 percent of pump power is reflected, and affects half of the spins probed.}
\end{figure}

To zero transverse magnetic fields, we adjust pump beam optics until \(\bf{s}_p\) points along \(B_z\). This is done far away from \(B^c\), such that electron spins are decoupled from nuclear spins. When aligned, transient signal amplitude \(\sqrt{A^2+B^2}\) should be zero. Due to presence of residual magnetic fields inside the magnetic shields, instead of minimizing oscillation amplitude at one \(B_z\) value when applied transverse magnetic fields are set to zero, the alignment is done by measuring and thereafter minimizing the slopes of transverse fields \(B_x\) and \(B_y\) that zeroes \(\sqrt{A^2+B^2}\) as functions of \(B_z\). Carrying out the same procedure around \(B^c\) allows us to zero light shift from probe laser with a stress plate. Since larger magnetic fields need to be applied to cancel the effective field from light shift closer to the compensation point due to compensation from nuclear spins, dispersive and absorptive shapes can be observed, as shown in Fig.~\ref{fig:dispersion}. At the compensation point an applied \(B_y\) cancels \(L_x\) due to imperfect self-compensation. We also verified pump light shift has no effect by checking that tuning pump laser wavelength results in little difference in such curves.

Examining the Bloch equations, one can see that when zeroing out \(\sqrt{A^2+B^2}\) with applied transverse magnetic fields, presence of any anomalous coupling would also require larger magnetic fields closer to the compensation point, resulting in similar dispersive and absorptive shapes. Typically other terms are small compared to \(\bf{L}\), with the next dominant one being Earth's rotation. By comparing \(B_{x/y}\) at the compensation point and away from it, one can differentiate field offset due to anomalous fields from residual magnetic fields, and get rid of the signal offsets in Fig.~\ref{fig:CPID}(b).

However, as Fig.~\ref{fig:dispersion} also shows, a rotation of the cell dramatically changes the shape of the curves, while effect from the probe beam remains small, which suggests imperfections not directly captured by Eqn.~\ref{eqn:Bloch}. Model-wise we were able to generate similar shapes by introducing non-uniform illumination of Rb spins by the pump beam. In the simplest case we use two Bloch equations to model Rb spins, uncoupled to each other but both coupled to the Bloch equation for Ne spins. The mean of the two Rb polarizations' \(\hat{x}-\)projection is taken as the comagnetometer signal. Similar shapes as seen in left panel of Fig.~\ref{fig:dispersion} arise under various scenarios, including when the two Rb ensembles are pumped along directions differing by a few degrees, when one ensemble sees an additional reflected beam with reduced intensity non-parallel to the incoming pump beam, or when the two ensembles have different average \(P^e_z\) and feel different transverse magnetic fields. All three scenarios can be linked to imperfections in the comagnetometer cell's optical properties. We confirmed with a cell from the same batch as the one used in experiment that the top window has different effective focal length across its surface and along different directions. Its cross-section also appears to be a wedge, thicker on one side than the other. Another piece of evidence suggesting non-uniform illumination is observed distortions in the experimental transient signal, making it deviate from the form specified by Eqn.~\ref{eqn:fittingForm}. This indicates that not all Rb spins probed share the same dynamics. 

\begin{figure}
\includegraphics[width=8.5cm, keepaspectratio]{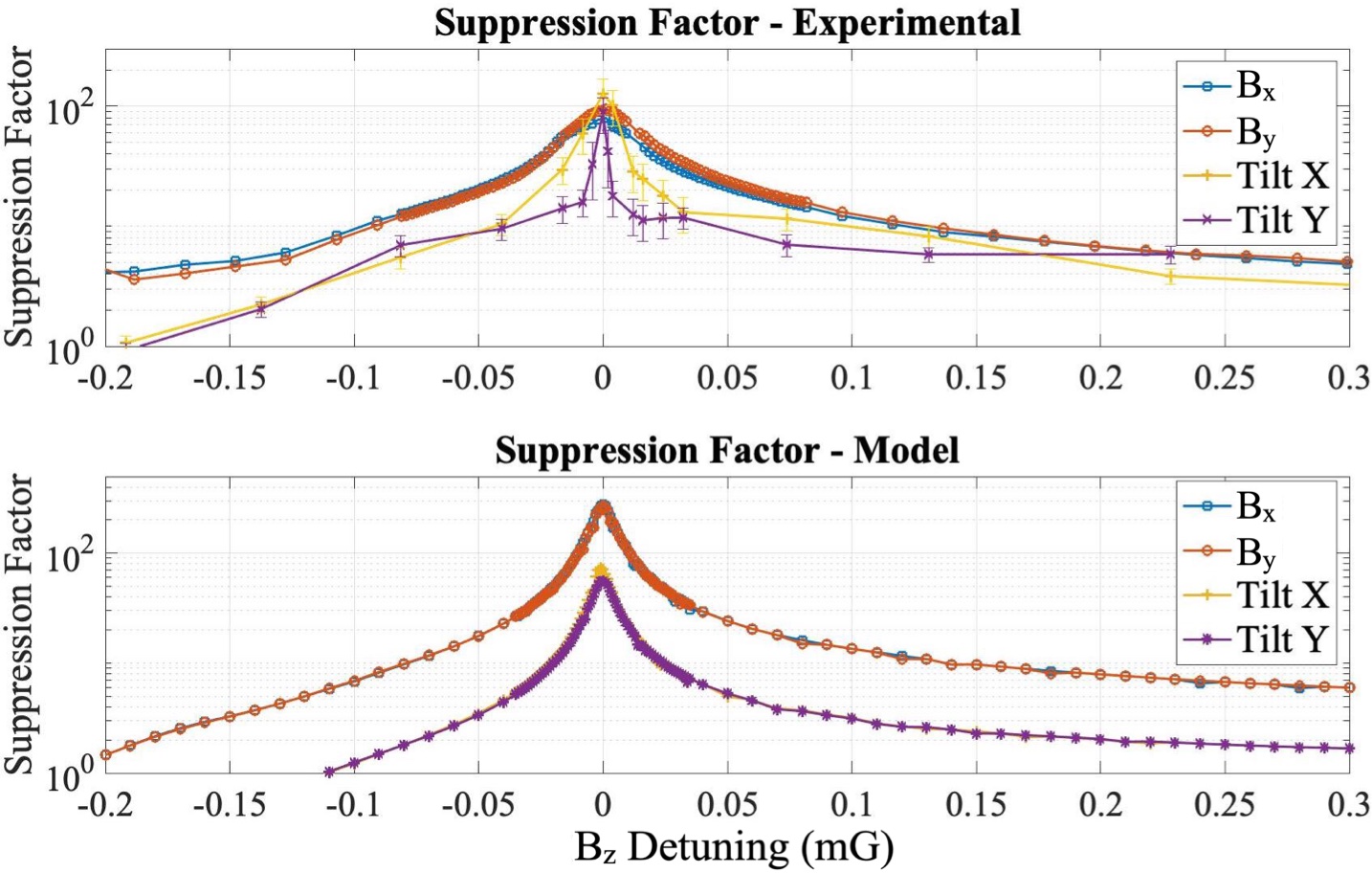}
\caption{\label{fig:SuppFactor} Suppression factor (see text for definition) of transverse magnetic field and pump beam deflection around the compensation point. Experimental curves for pump beam tilt into \(\hat{x}\)- and \(\hat{y}\)-axis do not completely overlap, suggesting that imperfections in the glass cell may result in non-uniform and asymmetrical illumination of alkali spins by the pump beam.}
\end{figure}

In Fig.~\ref{fig:SuppFactor} we demonstrate the pulsed comagnetometer's suppression of transverse magnetic fields and of pump beam deflections. The suppression factor here is defined as the ratio between signal response of a pulsed magnetometer where \(\bf{P^n}=0\) and that of the pulsed comagnetometer to the same transverse field or pump beam tilt. Detuning for the pulsed magnetometer is any deviation of applied \(B_z\) from \(-B^e\), so the electron spins experience the same longitudinal field at the same \(B_z\) detuning for both the magnetometer and the comagnetometer. Ne polarization can be maintained at zero by rapidly flipping the pump beam's circular polarization handedness with a liquid crystal variable waveplate. As noted before, good pump beam tilt suppression requires well-collimated beam such that the \(uvw-\)frame and the equivalent magnetic fields are well-defined. Variations in the beam's intensity profile is of lesser importance at high pumping rate. Experimentally we found that pump beam tilt suppression works best when the center of pump beam hits a specific point on top cell window, possibly because glass wall properties specific to the cell result in the most uniform illumination at that point. Care should also be taken to chase all Rb droplets into the stem to minimize reflection of the pump beam, instead of having droplets smeared across the cell's bottom surface.

\begin{figure}
\includegraphics[width=8.5cm, keepaspectratio]{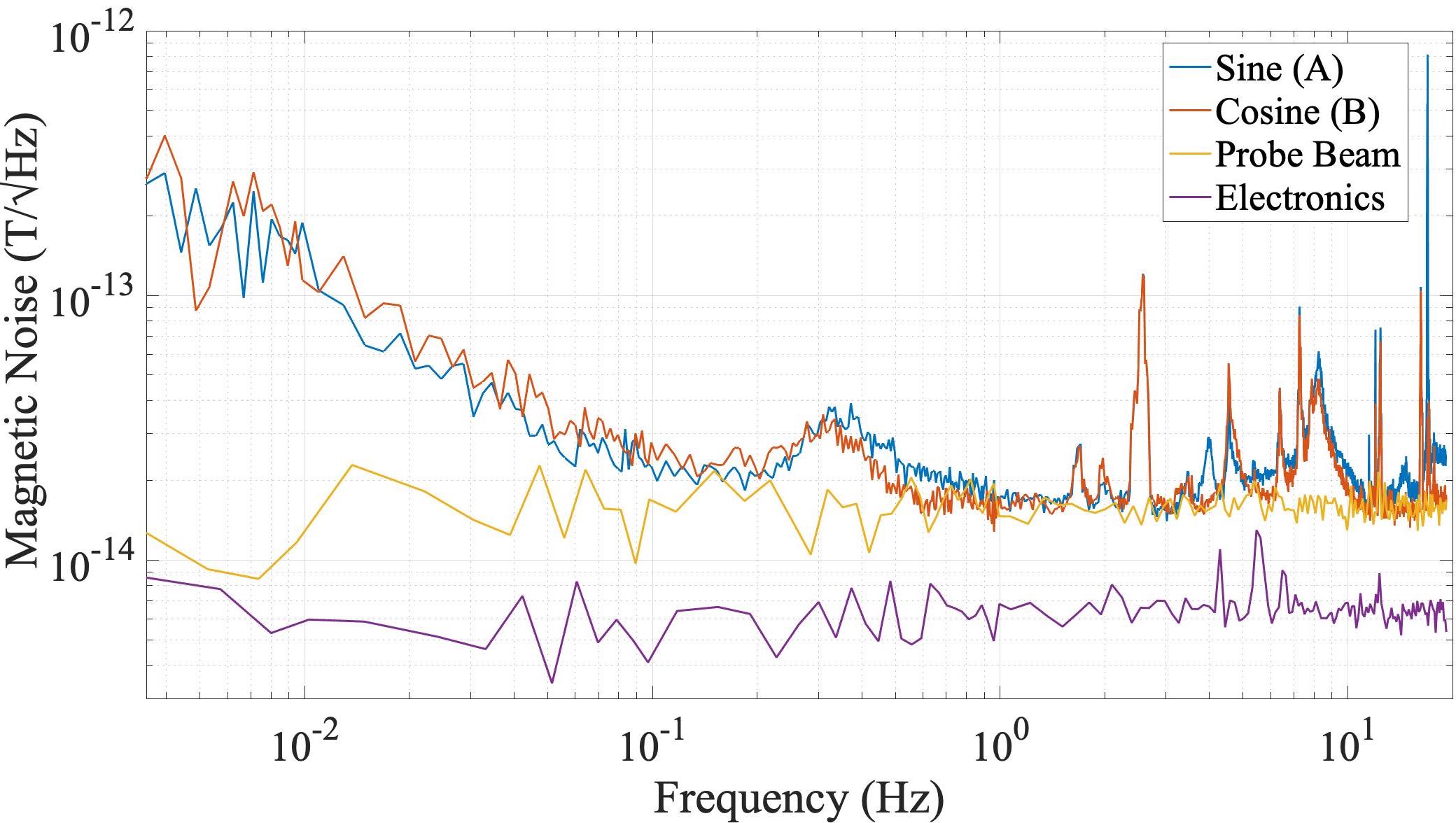}
\caption{\label{fig:FFT} Fourier spectrum of the pulsed comagnetometer noise. Probe beam curve is for fitted parameter \(B\) when pump laser is off, and electronics curve is also for \(B\) when both pump and probe lasers are off.}
\end{figure}

The sensitivity of the pulsed comagnetometer is shown in Fig.~\ref{fig:FFT}. 
Magnetic field sensitivity in the white noise region is about 20 fT\(/\sqrt{\text{Hz}}\), limited by photon shot noise. Higher short term sensitivity can be obtained by reducing the pump duration to reduce
the value of \(s\), and by using the data closer to the start of dark precession with a more elaborate model that includes variation of precession frequency or by using a template function. We estimate the sensitivity could be improved to 6.6 fT\(/\sqrt{\text{Hz}}\). The broad peak around 0.3 Hz is the resonance frequency of the coupled spins. The response of the co-magnetometer drops off above this frequency,
so the peaks at higher frequency are not relevant. It is desirable to lengthen Ne's lifetime such that this peak can be shifted to improve sensitivity in that frequency range. Currently Ne spin-relaxation is predominantly limited by the gradient of its dipolar field due to the aspherical cell geometry. The \(1/f\) noise knee stays at the same frequency of about 0.1 Hz compared to the DC-mode comagnetometer with the best sensitivity \cite{vasilakis09}. It is not yet shifted to lower frequency most likely due to the large dispersion and absorption amplitudes in Fig.~\ref{fig:dispersion}, which means there exists an unsuppressed effective pseudo-field in the transverse plane.

\section{\label{sec:level1}Conclusion}

We devise and realize a self-compensating alkali-metal-noble-gas comagnetometer with pulsed optical pumping, which improves upon previous generations of DC-mode self-compensating alkali-metal-noble-gas comagnetometers, while inheriting their inherent suppression of magnetic fields and high sensitivity from being in the near-SERF regime. The pulsed comagnetometer has dual-axis sensitivity, avoids pump laser light shifts by measuring in the dark, suppresses perturbation from pump and probe beam pointing fluctuations, and is insensitive to pump power fluctuations at sufficiently high pumping rate. We also derive an analytic model and describe unique features of the comagnetometer operation in the pulsed regime. A zeroing procedure is designed to minimize the remaining systematic effects, while concurrently calibrating the comagnetometer. With the above advantages, the pulsed comagnetometer exhibits good short-term sensitivity. 
While we anticipated improved \(1/f\) noise performance compared to the DC-mode of operation, we find that pump beam spatial inhomogeneity caused by vapor cell defects prevents us from realizing reduced \(1/f\) noise. Using anodically bonded cell with optimized wall curvature would allow more precise laser beam collimation and increase Ne's equilibrium polarization.

This work was supported by NSF and Simons Foundation.

\nocite{*}

\bibliography{RbNePulsedComag.bib}% Produces the bibliography via BibTeX.

\end{document}